\documentclass{article}

\usepackage{PRIMEarxiv}

\usepackage[utf8]{inputenc} 
\usepackage[T1]{fontenc}    
\usepackage{hyperref}       
\usepackage{url}            
\usepackage{booktabs}       
\usepackage{amsfonts}       
\usepackage{nicefrac}       
\usepackage{microtype}      
\usepackage{lipsum}
\usepackage{graphicx}
\graphicspath{{media/}}     
\usepackage{amsmath}
  
\title{Advancing Stepped Wedge Cluster Randomized Trials Analysis: Bayesian Hierarchical Penalized Spline Models for Immediate and Time-Varying Intervention Effects 
}

\author{
  Danni Wu \\
  Department of Population Health \\
  New York University Grossman School of Medicine \\
  New York\\
  \texttt{dw2625@nyu.edu} \\
   \And
Hyung G. Park \\
  Department of Population Health \\
  New York University Grossman School of Medicine \\
  New York\\
  \texttt{Hyung.Park@nyulangone.org} \\
   \AND
   Corita R Grudzen\\
   Department of Medicine\\
   Memorial Sloan Kettering Cancer Center\\
   New York\\
  \texttt{corita.grudzen@nyumc.org} \\
   \AND
   Keith S. Goldfeld\\
  Department of Population Health \\
  New York University Grossman School of Medicine \\
  New York\\
  \texttt{Keith.Goldfeld@nyulangone.org} \\
}

\begin{document}

\maketitle

\begin{abstract}
\noindent\textbf{Background:}
Stepped wedge cluster randomized trials (SWCRTs) are pivotal in healthcare to introduce interventions ethically and logistically across various clusters over time. These trials, while minimizing contamination risks between intervention and control groups, face challenges with potential confounding by temporal trends. Traditional frequentist methods, designed to adjust for confounding by temporal trends, can fail to provide adequate coverage of the intervention's true effect using confidence intervals,  whereas Bayesian approaches,  show potential for better coverage of intervention effects. However, Bayesian methods have seen very limited development in SWCRTs, suggesting a need for further research and refinement in this area.

\noindent\textbf{Methods:}
We propose two novel Bayesian hierarchical penalized spline models for SWCRTs. The first model is designed for SWCRTs involving a relatively large number of clusters and time periods, focusing on immediate intervention effects. It leverages the Bayesian hierarchical penalized spline to model the complex, nonlinear cluster-specific time trends. To evaluate its efficacy, we compared this model to traditional frequentist methods, focusing on metrics such as bias, root mean square error (RMSE), and coverage probability.
Building upon this, we further developed the model to estimate time-varying intervention effects. This extension employs Bayesian hierarchical penalized splines, offering a flexible approach to model temporal trends and complex patterns. We conducted a comparative analysis of this Bayesian spline model against an existing Bayesian monotone effect curve model, examining their bias (absolute value), RMSE, and leave-one-out cross-validation information criterion (LOOIC). The proposed models are applied in the ``Primary Palliative Care for Emergency Medicine (PRIM-ER)'' study, which uses a cluster-randomized, stepped wedge design to evaluate the effectiveness of primary palliative care through education, training, and technical support in the field of emergency medicine.

\noindent\textbf{Results:}
Extensive simulations and an application in the PRIM-ER study demonstrate the strengths of the proposed Bayesian models. The Bayesian immediate effect model consistently achieves near the frequentist nominal coverage probability for true intervention effect, providing more reliable interval estimations than traditional frequentist models, while maintaining low RMSE and bias. The proposed Bayesian spline model, designed to estimate time-varying intervention effects, exhibits advancements over the existing Bayesian monotone effect curve model in terms of improved accuracy and reliability.

\noindent\textbf{Conclusion:}
The proposed Bayesian hierarchical spline models offer a more accurate and robust analysis of intervention effects in SWCRTs. Their application could lead to more effective and timely adjustments in intervention strategies, ultimately enhancing patient outcomes.

\end{abstract}


\section{Background}
\label{sec:4_background}
The stepped wedge cluster randomized trial (SWCRT) is a type of cluster randomized trial in which the clusters (e.g.,   hospitals, public health units, or communities) are assigned in a random sequence to start receiving the intervention. Initially, all clusters begin with a control intervention. Once a cluster starts the intervention, all participants in that cluster continue to receive the intervention for the rest of the study period. During the course of the trial, every cluster will have transitioned from control to intervention \cite{hemming2015stepped, 10.1001/jama.2017.21993} so that by the end of the trial, all clusters are receiving the intervention. The strengths and limitations of the stepped wedge design have been a subject of extensive discussion in recent literature \cite{brown2006stepped,mdege2011systematic,kotz2012use,mdege2011systematic,kotz2012researchers,hemming2013stepped,hemming2015stepped,kotz2013stepped, beard2015stepped}. Notably, this design is particularly favored in scenarios (1) where simultaneous intervention implementation across multiple clusters is impractical due to logistical, financial, or other constraints \cite{brown2006stepped}; (2) when it is considered unethical to withhold an intervention from any cluster, as every cluster eventually receives the intervention in SWCRT \cite{beard2015stepped,mdege2011systematic,barker2016stepped}; (3) where there is a risk of contamination between treatment and control subjects; the SWCRT minimizes contamination by sequentially introducing the intervention to different clusters over time, reducing the likelihood of interaction and information exchange about the intervention between the control and intervention groups \cite{barker2016stepped, nickless2018mixed, 10.1001/jama.2017.21993}. However,  SWCRT may lead to confounding by temporal trends. Since changes in clinical care occur over time, comparisons of outcomes between earlier and later periods may be influenced by background changes that affect the outcome of interest irrespective of the intervention being tested \cite{10.1001/jama.2017.21993}. Limitations arise when simple models that ignore the effect of time are used, leading to potential confounding between the effect of time and the intervention. Complex temporal trends can impact outcomes, necessitating flexible and comprehensive modeling strategies \cite{nickless2018mixed}. 

\begin{figure}[htbp]
  \centering
\includegraphics[width=\textwidth]{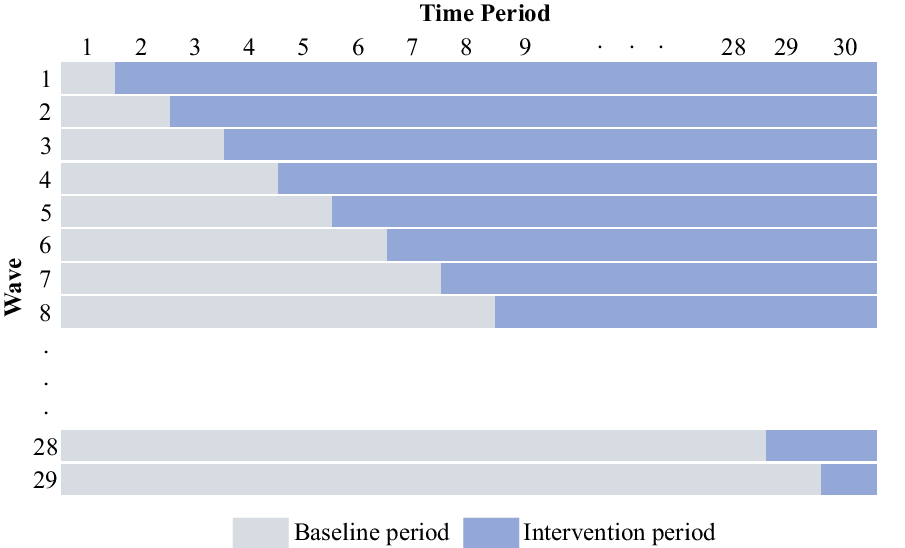}
   \caption{Design of the motivated stepped wedge cluster randomized trial with a relatively large number of waves and time periods: The plot depicts a stepped wedge cluster randomized trial with 29 clusters over 30 time periods. Rows represent clusters, and columns represent time periods. Lighter shades indicate the baseline control period, while darker shades show when each cluster receives the intervention. Clusters sequentially transition to the intervention, with all clusters participating by the final period.}
  \label{fig:29_waves}
\end{figure}

In the classic stepped wedge trial design, a small number of clusters are grouped into `waves' and are randomized to receive the intervention at different starting times \cite{grantham2022evaluating}. However, our study employs a unique variant of this design due to the complexity of the intervention, necessitating its sequential introduction across individual clusters. This design results in a scenario where each cluster forms its own wave, leading to a significantly increased number of waves. This increase in waves translates into a corresponding increase in the number of time periods over which the intervention is implemented, as an example shown in Figure \ref{fig:29_waves}. Consequently, we face the challenge of having to estimate a larger set of period-specific effects using conventional methods \cite{nickless2018mixed,kenny2022analysis}, while each time period, being shorter, provides less data to accurately determine these effects. This situation poses both a methodological and analytical challenge in evaluating the effect of the intervention across many, shorter time periods. To address these challenges, smoothing the time trend using spline models, especially penalized splines, appears promising. This approach, distinct from conventional methods that estimate separate time effect coefficients for each time period \cite{nickless2018mixed,kenny2022analysis}, enables flexible modeling of complex, non-linear temporal trend while significantly reducing the number of parameters requiring estimation \cite{suk2019nonlinear}. However, traditional methods like linear mixed-effects models \cite{bates2014fitting} or generalized additive models \cite{wood2017generalized}, estimated via maximum likelihood or restricted maximum likelihood (REML), have been shown to underestimate the intervention effect's 95\% confidence interval coverage for the true value \cite{ourdatageneration}. Bayesian approaches, known for their capacity to model complex hierarchical structures, ``borrow information" across clusters, and allow for probability-based inferences for any parameter of interest, emerging as a promising alternative \cite{Gelman2013BayesianAnalysis}. However, Bayesian models have had very limited application in SWCRTs \cite{cunanan2016practical, grantham2022evaluating,zhan2021improving,kenny2022analysis}. To the best of our knowledge, Bayesian spline \cite{lang2004bayesian} has not been developed for modeling time trends in a SWCRT. 

Besides the need for advanced Bayesian methods for modeling time trends, there exists another gap in the literature: the use of Bayesian spline modeling to address time-varying intervention effects. Conventional SWCRTs often rely on models assuming immediate intervention effects (i.e., as soon as the intervention starts, its full impact is realized suddenly), which is suitable in some scenarios, such as the impact of a surgical safety checklist on patient outcomes would be immediate \cite{haugen2015effect}. However, this assumption is less plausible in cases where effects, like those from education programs or community health initiatives, emerge gradually \cite{menon2016impacts, white2018community}. Recognizing the limitations of the immediate effect assumption, Granston \cite{granston2014addressing} initially tackled the challenge of time-varying intervention effects in SWCRTs by proposing a complex two-stage estimation process for concave effect curves, but this model can lead to incorrect inference when the effect curve does not follow the assumed distribution and can fail with non-concave curves. Hughes et al. \cite{hughes2015current} suggested study design adjustments and modifying the treatment indicator variable, but this assumes that the shape of the effect curve is fully known. Hemming et al. \cite{hemming2017analysis} introduced a model using study time-treatment interaction terms with study time treated as a categorical variable, but encountered issues with wide confidence intervals. Nickless et al. \cite{nickless2018mixed} and Kenny et al. \cite{kenny2022analysis} explored various models accounting for time-varying effects, revealing the risks of assuming immediate effects. Despite these efforts, the development of Bayesian spline modeling for estimating time-varying intervention effects in SWCRTs remains an unexplored area.

These identified gaps in the literature highlight a significant opportunity for advancing Bayesian methodological approaches in SWCRTs to improve the accuracy and reliability of intervention effect estimations. We propose two Bayesian hierarchical penalized spline models: the first model focuses on evaluating immediate intervention effects using Bayesian penalized spline for modeling time trends in SWCRTs with a relatively large number of clusters and time periods, aiming at improving coverage.
Building on the first model and recognizing that intervention effects can increase, decrease, or follow non-linear patterns throughout the exposure period, we extend the model to incorporate time-varying intervention effects, employing Bayesian penalized spline to capture the temporal dynamics of intervention effects. Such models are particularly crucial in the context of medical education and training, where understanding how the magnitude of the treatment effect changes with increasing experience in treatment is key to effective policy and practice \cite{kenny2022analysis,hopper2007learning}. By offering a better understanding of how intervention effects evolve over time, our models can provide a tool for assessing the effectiveness of various interventions.

We organize the chapter as follows. In \hyperref[sec:4_methods]{the Methods section}, we propose a Bayesian hierarchical spline model for evaluating immediate intervention effects. Additionally, we extended the Bayesian hierarchical spline model to accommodate time-varying intervention effects. Furthermore, we developed a Bayesian time-varying effect model that includes cluster-specific random effects, which allow for the intervention effect curve to vary between clusters. We elucidate the reasoning behind our selection of prior distributions. In \hyperref[sec:4_results]{the Results section}, 
we compare the proposed Bayesian models with existing approaches through extensive simulations. Furthermore, we have applied our model to data from a real-world SWCRT. In \hyperref[sec:discuss]{the Discussion and conclusions section}, we provide a discussion and offer insights into the potential future directions of our research.
\section{Methods}
\label{sec:4_methods}
In this section, we present Bayesian hierarchical penalized spline models for immediate and time-varying intervention effects in SWCRTs. Suppose that in a SWCRT, there are $T+1$ equally spaced time points with $T$ sequences in a SWCRT. The initial phase serves as a baseline period during which no clusters are subjected to the intervention. After this baseline period, in each subsequent sequence, one cluster is randomly chosen to transition from the control group to the intervention group. This process is repeated until all of the clusters are receiving the intervention. Following the sequence in which the last cluster receives the intervention, an additional sequence is conducted to perform a follow-up on all of the clusters. For the models proposed here, it is assumed that measurements of individual patient outcomes are cross-sectional in nature.  These measurements are collected for each of the clusters at each of the time periods. 

Here, we introduce a model for a data-generating mechanism. Let $Y_{ijt}$ represent the observed outcome for an individual $i \in \{1, \ldots, I \}$ within cluster $j \in \{1,\ldots, J\}$ at time $t \in \{0, \ldots , T\}$.  $Y_{ijt}$ follows an exponential family distribution and $g(.)$ is the canonical link function depending on the distribution of outcome (e.g., identity function for continuous outcomes, logit function for binary outcomes, and log function for count outcomes). Our model is designed for the stepped wedge design when there are many clusters and time periods. This allows us to treat time as a continuous variable. The intervention status for cluster \( j \) at time \( t \) is denoted by \( A_{jt} \). Specifically, \( A_{jt} = 1 \) indicates that cluster \( j \) is under the intervention at time \( t \); otherwise,$A_{jt}$ = 0.
The data are generated as follows:
    \[g(\mathbb{E}[Y_{ijt}|t,A_{jt}])= \alpha_j + s_j(t) + \tau \cdot A_{jt}, \]
where $\alpha_j$ represents cluster-specific random intercept, $s_j(t)$ represents cluster-specific temporal effect, $\tau$ represents the intervention effect. When modeling the immediate effect of an intervention, which implies that the intervention reaches its full effect instantly after initiation and remains constant, $\tau$ is a parameter not associated with time. However, when modeling the time-varying intervention effects, which implies that the intervention’s effect could change over time as the cluster adapts to the
intervention, $\tau$ is considered as a function of the time variable.

\subsection{Bayesian hierarchical penalized spline model for immediate intervention effects}
\label{sec:fixed_trt_model}
We propose a Bayesian hierarchical penalized spline model, which can model the relationship between the covariates and the outcome variable while simultaneously incorporating penalization terms to promote smoothness in the spline function. 
The model is set up as:

\begin{align} 
  \begin{split}
g(\mathbb{E}[Y_{ijt}|t,A_{jt}])&= \alpha  + s_j(t) + \tau \cdot A_{jt}\\ 
  &=\alpha  + \boldsymbol{\beta}_{bj} \cdot \mathbf{B} + \tau \cdot A_{jt}\\
\alpha & \sim \text{Normal}(0, 1) \\
\boldsymbol{\beta}_{bj} & \sim \text{Normal}(\boldsymbol{\beta}, \sigma_b^2I_{p \times p})\\
\boldsymbol{\beta} &= (\beta_{1},  \ldots, \beta_{p})\\
\beta_{1}& \sim \text{Normal}(0, 1)\\
\beta_{m}& \sim \text{Normal}(\beta_{m-1}, \sigma_{\beta}^2),\quad m=2,\ldots,p\\
\sigma_b &\sim t_\text{student}(df=3, \mu=0, \sigma=2.5)\\
\sigma_{\beta} &\sim \text{Normal}(0, 1)\\
\tau & \sim \text{Normal}(0, 5^2) 
\label{eq:Bayes}
\end{split}
\end{align} 

\begin{itemize}
\item $Y_{ijt}$ is the response variable with link function $g(.)$.
\item $A_{jt}$ is a binary intervention status for cluster $j$ at time point $t$.
\item $\alpha$ is the overall intercept.
\item $\boldsymbol{B} \in \mathbb{R}^{p \times 1} $ represents B-spline basis vector for time $t$. In our implementation, B-spline basis functions are generated from the time variable $t$. The knots, which determine where the polynomial pieces join, are determined based on quantiles of the total study time. Specifically, six quantiles are selected, excluding the smallest and largest quantiles to ensure that the splines are encouraged to be smoother near the edges of the data range, often providing a more stable prediction at the edges. The degree of the splines, which specifies the degree of the polynomial pieces, is set to three, making them cubic B-splines.
We tested using seven and ten knots and found no significant change in accuracy, but more knots increased computational time.
\item $\boldsymbol{\beta}_{bj}$ is the vector of coefficients associated with B-spline basis functions in cluster $j$. We used the Bayesian hierarchical modeling technique for $\boldsymbol{\beta}_{bj}$. The $\boldsymbol{\beta}_{bj}$'s have a common mean; the variation is $\sigma_b$. This prior distribution assumes that each vector of cluster-specific spline coefficient $\boldsymbol{\beta}_{bj}$  is centered around a vector of pooled coefficient $\boldsymbol{\beta}$, postulating that these cluster-specific vectors are correlated across all clusters. The prior distribution of $\sigma_b$ is $t_\text{student}(df=3, \mu=0, \sigma=2.5)$, facilitating ``borrowing of information'' from the clusters while allowing for large variability across clusters.

\item $\boldsymbol{\beta}$ represents a vector of pooled coefficients corresponding to B-spline basis functions across all clusters, and it has a dimension of $p$. We used a random-walk prior for $\boldsymbol{\beta}$, enforcing smoothness across the coefficients. The idea behind the random walk is that if the coefficients of nearby B-splines are close to each other, we will have less local variability\cite{lang2004bayesian, randomwalk}. This particular choice of a random-walk prior not only enforces smoothness across the coefficients but also serves as a mechanism to penalize overfitting \cite{lang2004bayesian}. 
\item $\tau$ represents the intervention effect across clusters.
\item When dealing with a Gaussian outcome, its dispersion (noise standard deviation) parameter $\sigma_{\epsilon}$ is assigned with a weakly informative prior, $\sigma_{\epsilon} \sim t_\text{student}(df=3, \mu=0, \sigma=2.5)$.

\end{itemize}

 We also proposed another way to promote smoothness in the spline function, the model adds a penalty to the log density for large differences between consecutive B-spline coefficients. This penalty is imposed on the second-order difference of the spline coefficients (i.e., $0.5\lambda(\beta_{m-1}-\beta_{m}+\beta_{m+1})^2, m=2,\ldots, p-1$, in which $p$ is the length of $\boldsymbol{\beta}$), which effectively forces adjacent coefficients to be similar, thus making the spline curve smoother. The hyperparameter $\lambda$ controls the strength of this penalization. The prior distribution for $\lambda$ reflects our belief about how much those second-order differences should be penalized in the model. We use a weakly informative prior distribution $t_{\text{student}}(df=3,\mu=0,\sigma=2.5)\cdot \mathbb{I}(\lambda \geq 0)$ for $\lambda$, where $\mathbb{I}$ is an indicator function. This prior distribution suggests that while we have some belief about the level of smoothness, we are still allowing the data to inform the final model. 

By incorporating B-splines into a Bayesian framework, we harness the flexibility of splines while benefiting from the probabilistic interpretation and regularization offered by the Bayesian paradigm. The model provides a robust and flexible approach to capture non-linear trends in data while avoiding overfitting through the incorporation of penalization terms.

\subsection{Bayesian hierarchical penalized spline model for time-varying intervention effects}
\label{sec:varying_trt_model}
The previous section is formulated under the assumption that the intervention's full effect is reached immediately after its initiation and remains constant. However, this may not always be true. In SWCRTs, each cluster begins the treatment at a different time and is then observed across multiple time periods. It's possible that the strength of the intervention effects could change over the duration of intervention as the cluster `learns' or adapts to the intervention — a phenomenon commonly observed in fields like surgery \cite{hopper2007learning}, known as the ``learning curve''. Therefore, such models in Section \ref{sec:fixed_trt_model} may not reflect the time-varying nature of real-world interventions.

To account for the variability of intervention effects over time, we extended the Bayesian immediate effect model \eqref{eq:Bayes} to model the intervention effect as a function of exposure time. The use of ``exposure time'' aligns with the approaches of Nickless et al. \cite{nickless2018mixed} and Kenny \cite{kenny2022analysis}. These authors suggest that the intervention effects may vary as a function of exposure time. Exposure time is defined as the duration since the initiation of the intervention in a given cluster, starting from zero at the beginning of the intervention. This concept is distinct from ``study time'', which refers to the elapsed time since the onset of the entire study. By incorporating this distinction, the proposed Bayesian time-varying effect model can more accurately capture the intervention effect as it evolves over the period of a cluster's exposure to the intervention. 

The proposed Bayesian hierarchical penalized spline model for time-varying intervention effects is set up as follows:

\begin{align} 
  \begin{split}
 g(\mathbb{E}[Y_{ijt}|t, t^{*}_j, A_{jt}])&= \alpha  + s_j(t) + \tau(t^{*}_j) \cdot A_{jt}\\ 
  &=\alpha  + \boldsymbol{\beta}_{bj} \cdot \mathbf{B} + \boldsymbol{\beta}^{*} \cdot \mathbf{B^{*}}\cdot A_{jt} \\ 
\alpha & \sim \text{Normal}(0, 1) \\
\boldsymbol{\beta}_{bj} & \sim \text{Normal}(\boldsymbol{\beta}, \sigma_b^2I_{p \times p})\\
\boldsymbol{\beta} &= (\beta_{1},  \ldots, \beta_{p})\\
\beta_{1}& \sim \text{Normal}(0, 1)\\
\beta_{m}& \sim \text{Normal}(\beta_{m-1}, \sigma_{\beta}^2),\quad m=2,\ldots,p\\
\sigma_{b} &\sim t_\text{student}(df=3, \mu=0, \sigma=2.5)\\
\sigma_{\beta} &\sim \text{Normal}(0, 1)\\
\boldsymbol{\beta}^{*} &= (\beta^{*}_{1},  \ldots, \beta^{*}_{p})\\
\beta^{*}_{1} & \sim \text{Normal}(0, 1) \\
\beta^{*}_{m}& \sim \text{Normal}(\beta^{*}_{m-1}, \sigma_{\beta^{*}}^2),\quad m=2,\ldots,p\\
\sigma_{\beta^{*}} & \sim \text{Normal}(0, 1)
\label{eq:4_Bayes_interaction}
 \end{split}
\end{align}

The notation largely follows model \eqref{eq:Bayes}.  The key change involves transforming the variable $\tau$ in model \eqref{eq:Bayes} into a function of exposure time, $\tau(t^{*}_j)$, where $t^{*}_j$ represents the exposure time of cluster $j$ at study time $t$, and $\tau(t^{*}_j)$ denotes the time-varying intervention effects. To effectively estimate this flexible function, we have employed Bayesian penalized B-splines. In this context, $\mathbf{B^{*}}$ represents B-spline basis vector for the exposure time, featuring $p$ knots. These B-spline basis functions are derived from the exposure time variable $t^{*}_j$.
The selection of knots for these splines is based on the quantiles of $max(t^{*}_j)$, specifically choosing six quantiles while excluding the smallest and largest to promote smoother spline behavior near the data range's edges. We opted for cubic B-splines by setting the splines' degree of the polynomial pieces to three. $\boldsymbol{\beta^{*}}$ represents a vector of coefficients corresponding to B-spline basis $\mathbf{B^{*}}$, with its dimension being $p$. To ensure smoothness across these coefficients, we utilized a random walk prior for $\boldsymbol{\beta^{*}}$. This random-walk prior ensures smoothness and penalizes overfitting, thus maintaining a balance in the model between accuracy and complexity.

\subsubsection{Incorporating cluster-specific random effect for time-varying intervention effects}
The previously proposed models assume that intervention effects do not vary across clusters, which may overlook potential cluster-specific variations in the intervention effects. This section explores extensions that incorporate cluster-specific time-varying intervention effects which allow for the intervention effect curves to vary across different clusters. We assume that the intervention effect curves specific to each cluster should fluctuate around an overall intervention effect curve. This overall curve is our primary focus, as it represents the pooled intervention effect across all clusters. The specification given in Model \eqref{eq:4_Bayes_interaction_randomeff} allows for the effect curve to vary across clusters and involves two additional parameters $u_j$ and $\sigma_{u_j}$:

\begin{align} 
  \begin{split}
 g(\mathbb{E}[Y_{ijt}|t, t^{*}_j, A_{jt}]) &= \alpha  + s_j(t) + \tau_j(t^{*}_j) \cdot A_{jt} \\ 
  &= \alpha  + s_j(t) + \tau(t^{*}_j)\cdot exp(u_j) \cdot A_{jt}\\ 
  &=\alpha  + \boldsymbol{\beta}_{bj} \cdot \mathbf{B} + \boldsymbol{\beta}^{*} \cdot \mathbf{B^{*}} \cdot exp(u_j)\cdot A_{jt}\\ 
  u_j &\sim \text{Normal}(0, \sigma_{u}^2)\\
\sigma_{u} &\sim \text{Normal}(0, 0.2^2)\\
\alpha & \sim \text{Normal}(0, 1) \\
\boldsymbol{\beta}_{bj} & \sim \text{Normal}(\boldsymbol{\beta}, \sigma_b^2I_{p \times p})\\
\boldsymbol{\beta} &= (\beta_{1},  \ldots, \beta_{p})\\
\beta_{1}& \sim \text{Normal}(0, 1)\\
\beta_{m}& \sim \text{Normal}(\beta_{m-1}, \sigma_{\beta}^2),\quad m=2,\ldots,p\\
\sigma_{b} &\sim t_\text{student}(df=3, \mu=0, \sigma=2.5)\\
\sigma_{\beta} &\sim \text{Normal}(0, 1)\\
\boldsymbol{\beta}^{*} &= (\beta^{*}_{1},  \ldots, \beta^{*}_{p})\\
\beta^{*}_{1} & \sim \text{Normal}(0, 1) \\
\beta^{*}_{m}& \sim \text{Normal}(\beta^{*}_{m-1}, \sigma_{\beta^{*}}^2),\quad m=2,\ldots,p\\
\sigma_{\beta^{*}} & \sim \text{Normal}(0, 1)
\label{eq:4_Bayes_interaction_randomeff}
 \end{split}
\end{align} 

The cluster-specific factor $\exp(u_j)$ represents the cluster-specific effect modifier for the $j$th cluster, which is used to account for the unique, cluster-specific variations in the intervention effect ($\tau(t^{*}_j)$). $u_j$ is assumed to follow a normal distribution with a mean of 0 and a standard deviation of $\sigma_{u}$, which is a hyperparameter that controls the magnitude of variation allowed across different clusters. $\sigma_{u}$ has a prior distribution Normal($0,0.2^2$). This reflects the prior belief that the cluster-specific intervention effect curves should generally align closely with the overall effect curve, while still allowing flexibility for the variations to be estimated from the data.

\section{Results}
\label{sec:4_results}

In this section, we conduct a comparative analysis between our proposed Bayesian models and the existing conventional methods across various study scenarios. Specifically, we compare the performance of the Bayesian immediate effect model, as outlined in Equation \eqref{eq:Bayes}, with that of traditional frequentist models. Additionally, we will examine the proposed Bayesian time-varying effect models \eqref{eq:4_Bayes_interaction} and \eqref{eq:4_Bayes_interaction_randomeff} alongside an existing Bayesian model that estimates a monotone effect curve. This comparison aims to evaluate their robustness, flexibility, and accuracy in estimating the intervention effects. Finally, we apply our Bayesian models to a SWCRT to demonstrate the practical applicability. Although we demonstrate the applicability and utility of our proposed models by using a continuous outcome in the simulation and a binary outcome in a real-world application as examples, the model is designed to be adaptable to other outcome types.

\subsection{Simulation illustration for Bayesian immediate effect model}

This section presents a simulation illustration of the proposed Bayesian model \eqref{eq:Bayes}. Additionally, it includes a comparative analysis of this model's performance against traditional frequentist approaches in SWCRTs.

\subsubsection{Simulation setup - Bayesian immediate effect model}
\label{sec:simuletion_setup_immediate}
We explore various scenarios by changing the intervention's true effect $\tau \in \{0,1,2,3,4,5\}$ during data generation, considering parameters $T=12$, $J=10$, $I=10$, and assuming a continuous outcome. The data is generated as follows:
\begin{enumerate}
    \item \textbf{Outcome}:
    \[Y_{ijt}= \alpha_j + s_j(t) + \tau \cdot  A_{jt} + \epsilon_{ijt} \]
    \item \textbf{Cluster-specific intercept}:
    \[ \alpha_j \sim \text{Normal}(u=0, \sigma=0.5) \].
    
        \item \textbf{Cluster-specific temporal effect}:
    \[ s_j(t) =  -0.01 t^2 + \gamma_{jt}, \]
    where $\gamma_{jt}$ is a cluster-specific time effect; the vector of cluster-specific time effects $\Gamma_j \sim    \text{MVN}(\boldsymbol{0},\Sigma_{T\times T})$, $\Sigma_{T\times T}=DRD$ is a $T\times T$ covariance matrix based on a diagonal matrix and an auto-regressive correlation structure $R$:
    \[D = 0.6\cdot I_{T\times T}\] and
    \[R = \begin{bmatrix}
1 & \rho & \rho^2 & \dots & \rho^{T-1} \\
\rho & 1 & \rho & \dots & \rho^{T-2} \\
\rho^2 & \rho & 1 & \dots & \rho^{T-3} \\
\vdots & \vdots & \vdots & \ddots & \vdots \\
\rho^{T-1} & \rho^{T-2} & \rho^{T-3} & \dots & 1 \\
\end{bmatrix}, \quad \rho = 0.95\]

 \item \textbf{Individual and time-specific noise} \( \epsilon_{ijt} \sim \text{Normal}(\mu=0, \sigma=1) \)
\end{enumerate}

In each scenario, the performance of the Bayesian model \eqref{eq:Bayes} is compared against commonly used frequentist models. 

\subsubsection{Overview of conventional frequentist models for immediate intervention effects}

For comparison, we employed four modeling strategies using the frequentist estimation paradigm. These models allow us to estimate the intervention effects while considering various structures of the time variable and cluster-level heterogeneity.

The initial model is a linear mixed effects model, utilizing the `lmer' function from the lme4 R package \cite{bates2014fitting}, which is saturated with respect to time. It allows us to account for the effect of time as a factor (with separate coefficients for each time point) and includes a random intercept for clusters and a random effect for time within each cluster.

\begin{align} 
  \begin{split}
Y_{ijt} = \alpha + \tau\cdot A_{jt} + \gamma_t  +  b_{1jt} + b_{0j} +\epsilon_{ijt}
\label{eq:linar_cate_time}
\end{split}
\end{align} 

where \( Y_{ijt} \) is the outcome for the \( i \)-th individual in cluster \( j \) at time \( t \), \(\alpha\) is the intercept, \( A_{jt} \) is the intervention status (equals to 0 or 1), \(\tau\) is the intervention effect, \( \gamma_t \) is the fixed effect of time \( t \), \( b_{0j} \) is the random intercept for cluster \( j \), \( b_{1jt} \) is the random effect for time \( t \) within cluster \( j \), and \( \varepsilon_{ijt} \) is the individual and time-specific noise. This model allows for different levels of \( t \) to have different impacts (i.e., it treats time as categorical with a fixed effect for each time point). This model is saturated with respect to study time, meaning it does not impose any specific assumptions on the shape of the time effect curve, assuming that measurements are taken at regular intervals. However, when applied to a stepped wedge design involving many time points and clusters, using this model might result in over-parametrization and computational burden. 

To examine the influence of treating time as a continuous variable, we utilized a linear mixed-effects model with time as a continuous variable and random slopes for time within clusters:
\begin{align} 
  \begin{split}
  Y_{ijt} = \alpha + \tau \cdot A_{jt} + (\beta_2 +  b_{1j}) \cdot t + b_{0j} + \varepsilon_{ijt} 
\label{eq:linar_conti_time}
\end{split}
\end{align} 
where \( \beta_2 \) is the fixed effect of time as a continuous variable, \( b_{0j} \) is the random intercept for cluster \( j \), \( b_{1j} \) represents the random slope for time within cluster \( j \).

In addition, the simplified linear model without the time component is considered for comparison; we focus solely on estimating the effect of the treatment while assuming the independence of observations and ignoring any changes over time. This model is specified as:
\begin{align} 
  \begin{split}
  Y_{ijt} = \alpha + \tau \cdot A_{jt} + \epsilon_{ijt} 
  \label{eq:linar_no_time}
\end{split}
\end{align} 
Additionally, we explored generalized additive models (GAMs) \cite{wood2017generalized} for a more flexible approach to modeling the continuous time effect, which includes a smooth term for the time trend in each cluster:
\begin{align} 
  \begin{split}
  Y_{ijt} = \alpha + \tau \cdot A_{jt} + s(t) + s_j(t) + \varepsilon_{ijt} 
\label{eq:GAM_time}
\end{split}
\end{align} 
We used the `bam' function from the `mgcv'  R package \cite{wood2015package}.
The \( s \) notation in the GAM models denotes the smooth functions used to model the non-linear relationships. \( s(t) \) is the smooth term for time, which captures the overall trend of time that is common to all clusters. \( s_j(t) \) represents a cluster-specific random smooth function for time, which captures the deviations from the overall time trend for each cluster. In the `mgcv' package, we use factor-smooth interactions, as indicated by the command `s(time, cluster, bs = ``fs'')'. This allows the model to fit distinct smooth curves for the time trend at each cluster level.

\subsubsection{Comparative evaluation of immediate effect models}

The metrics for comparison are (1) the percentage of simulations where the intervention's true effect falls within the estimated confidence intervals (CIs) or credible intervals (CrIs) (coverage of true value), (2) the root mean squared error (RMSE) between the estimated and the intervention's true effects, and (3) the bias (absolute value) between the estimated and the intervention's true effects. For Bayesian models, we calculated bias and RMSE using the posterior median as the point estimator of the intervention effect. The data generation process for simulations used the `simstudy'  \cite{Goldfeld2020PackageDocumented} R package, and model fitting uses Stan\cite{StanDevelopmentTeam2020} and R \cite{RCoreTeam2020}.  

\begin{figure}[h]
  \centering
\includegraphics[]{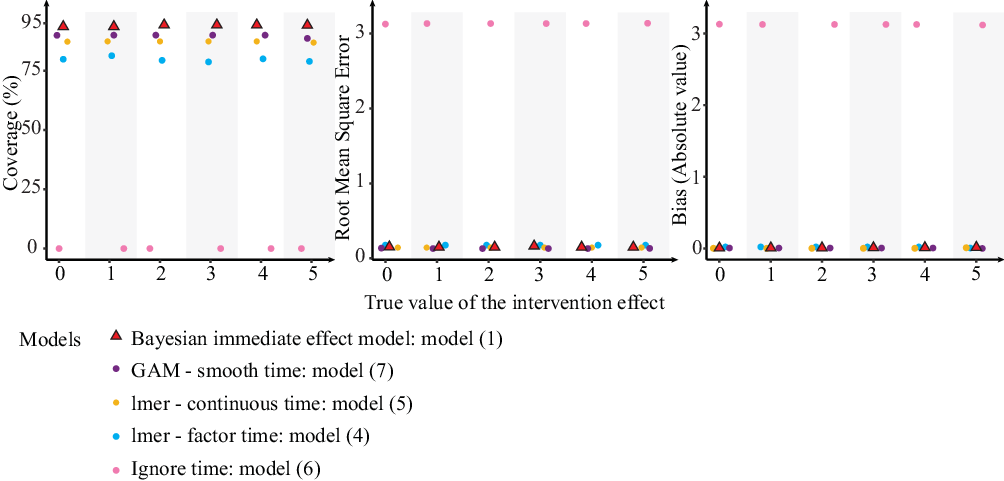}
  \caption{Comparative performance of the proposed Bayesian immediate effect model (1) and commonly used frequentist models (Model (4): linear mixed effect model treats time as a categorical variable; Model (5): linear mixed effect model treats time as a continuous variable; Model (6): linear model without time variable; Model (7): generalized additive model uses spline to model continuous time) in estimating the true effect of an intervention, illustrated across three metrics: coverage (left), root mean square error (RMSE, center), and bias (right). The x-axis represents the varying levels of the intervention's true effect in the data generation, while the y-axis shows the respective metric values for each model.}
  \label{fig:fixed_performance}
\end{figure}

As shown in Figure \ref{fig:fixed_performance}, the proposed Bayesian spline model \eqref{eq:Bayes} consistently demonstrates higher coverage probability across all scenarios, indicating a more reliable interval estimation compared to the frequentist models (models \eqref{eq:linar_cate_time} — \eqref{eq:GAM_time}). Specifically, the Bayesian models maintain a coverage close to 95\%, indicative of their robustness in encompassing the true parameter value. In contrast, the frequentist models exhibit variable coverage: for instance, model \eqref{eq:linar_cate_time} demonstrates approximately 79\% coverage, whereas model \eqref{eq:linar_no_time} fails to attain any coverage of the true value, exhibiting 0\% coverage.

The RMSE and bias results indicate that the accuracy of the proposed Bayesian spline model \eqref{eq:Bayes} is comparable to that of the frequentist models \eqref{eq:linar_cate_time}, \eqref{eq:linar_conti_time}, and \eqref{eq:GAM_time}, suggesting that both approaches yield estimates of the intervention's true effect with similar precision. Model \eqref{eq:linar_no_time}, which excludes the time effect, shows the worst performance in terms of RMSE and bias. This indicates a significant loss of accuracy and underscores the risks associated with not controlling for time confounding in SWCRTs.

\clearpage

\subsection{Simulation illustration for Bayesian time-varying effect model}

In this section, we assess the Bayesian time-varying effect model's performance through a series of simulation studies. This involved comparing our proposed Bayesian spline model \eqref{eq:4_Bayes_interaction} to the existing Bayesian monotone effect curve model \cite{kenny2022analysis} in terms of intervention effect estimation and model reliability.

\subsubsection{Simulation setup - Bayesian time-varying effect model}
\begin{figure}[p]
  \centering
\includegraphics[width= \textwidth]{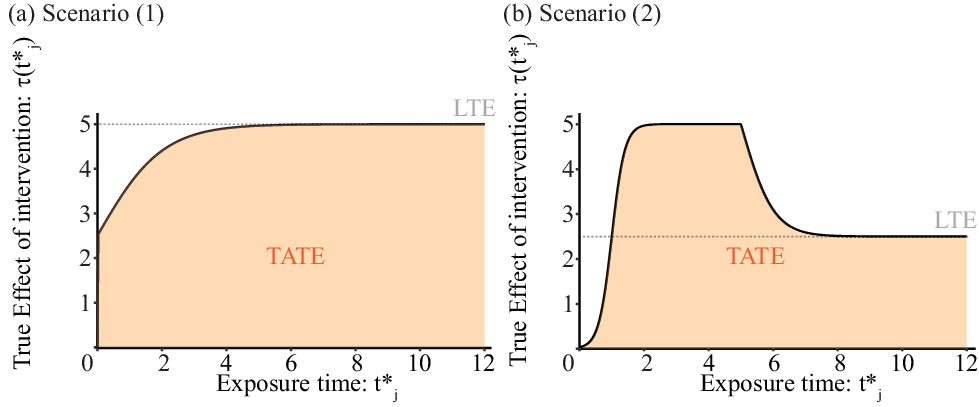}
\caption{Scenarios of generating time-varying intervention effects.  Two estimands are used to summarize the magnitude of the intervention effect curve: the ``time-averaged treatment
effect" (TATE) and the ``long-term treatment effect" (LTE).  TATE is calculated by dividing the area under the time-varying intervention effect curve by the total length of the exposure time. This measure represents the average effect of the intervention per unit of time throughout the study's duration. LTE captures the effect of the intervention at the study's longest exposure time point, focusing on long-term outcomes.}
\label{fig:vary_generate}
\end{figure}

The simulation scenarios, motivated by the use of education and training programs as interventions described in \hyperref[sec:4_application]{the Application section}, consider the following two scenarios where the intervention effect varies based on exposure time as suggested by domain experts. Figure \ref{fig:vary_generate} shows scenario (1) assumes a gradual increase in the intervention effects over time, eventually reaching its full potential. This gradual increase is based on the understanding that the impact of education and training may take months to materialize, as behavior change is often a slow process \cite{menon2016impacts,white2018community}. In Scenario (2), the intervention effect initially increases with the commencement of the training program on-site. This rise is due to the immediate application of new skills and knowledge by the trained personnel, fostering a productive and efficient work environment. As the training progresses, its effectiveness reaches a peak and stabilizes, indicating the successful adoption of the skills across the team. However, after reaching its peak, the effectiveness declines due to some organizational factors. Firstly, the departure of the training personnel results in a loss of ongoing guidance and mentorship, which is critical for reinforcing learned skills. Additionally, trained staff, now skilled, might leave for new roles, taking their expertise with them. Moreover, the arrival of new hires who have not undergone the training introduces a knowledge gap. These new team members may not be immediately aligned with the established methods and practices, thereby reducing the intervention's overall impact.
\begin{itemize}
 \item Scenario (1) (Figure \ref{fig:vary_generate}(a)): 
    \[
    \begin{cases} 
    0 & \text{if } t^{*}_j = 0 \\
    \frac{5}{1 + 2e^{-t^{*}_j}} & \text{if } t^{*}_j > 0 
    \end{cases}
    \]

 \item   Scenario (2) (Figure  \ref{fig:vary_generate}(b)):
    \[
    \tau(t^{*}_j) = 
    \begin{cases} 
    0 & t^{*}_j=0\\
    \frac{5}{1 + e^{-5 \times (t - 1)}} & \text{if } 5 > t^{*}_j > 0 \\
    2.5 + \frac{5}{1 + e^{2 \times (t - 5)}} & \text{if } t^{*}_j \geq 5 
    \end{cases}
    \]
\end{itemize}

We assume the outcome is continuous in the simulation: 

    \[ Y_{ijt} = \alpha_j + s_j(t) + \tau(t^{*}_j) \cdot A_{jt} + \epsilon_{ijt} \]

$\alpha_j$, $s_j(t)$, and $\epsilon_{ijt}$ is generated consistent with the setting outlined in Section \ref{sec:simuletion_setup_immediate}. Here, we set the number of clusters, denoted as $J$, to 10, and each cluster contains 10 participants, represented by $I=10$. 
We vary the number of time periods in these two scenarios:  in Scenario 1, there are 17 time periods ($T_{scenario 1} = 17$), while in Scenario 2, there are 22 ($T_{scenario 2} = 22$). For the first time sequence, all clusters were under control conditions; a cluster was randomly assigned to begin receiving the intervention at the start of the second time sequence. The last cluster started to receive the intervention at the beginning of time point 11, after which all clusters were followed up for the remainder of the study.

\clearpage

\subsubsection{Performance evaluation for Bayesian time-varying effect model}

In this section, we evaluate the performance of the proposed Bayesian spline model (referenced in Equation \ref{eq:4_Bayes_interaction}) in estimating the intervention effect curve. We used two estimands to summarize the magnitude of the intervention effect curve: the ``time-averaged treatment
effect'' (TATE) and the ``long-term treatment effect'' (LTE) \cite{kenny2022analysis}. 

The first
summary estimand is the TATE from the entire exposure period, spanning from study time $[2, T]$ (i.e., exposure time $t^{*} \in (0, T-2]$, so $t^{*}_{max}= T-2$), the TATE is defined as:
\[
\Psi_{(0,t^{*}_{max}]} = \frac{1}{t^{*}_{max}} \int_{0}^{t^{*}_{max}} \tau(t^{*}) \, dt^{*} = \frac{1}{T - 2} \int_{0}^{T-2} \tau(t^{*}) \, dt^{*}.
\]

where $\tau(t^{*})$ represents the effect curve, quantifying the intervention effect as a function of the exposure time $t^{*}$. We used the right-hand Riemann sum to approximate the integral of the time-varying intervention effect as suggested by Kenny\cite{kenny2022analysis}:

\[
\hat{\Psi}_{(0,t^{*}_{max}]} = \frac{1}{t^{*}_{max}} \sum_{t^{*}=1}^{t^{*}_{max}} \hat{\tau}(t^{*})=\frac{1}{T - 2} \sum_{t^{*}=1}^{T-2} \hat{\tau}(t^{*}) = \frac{1}{T - 2} \sum_{t^{*}=1}^{T-2} \sum_{m=1}^{p}\hat{\beta}^{*}_mb_m(t^{*}),
\]

where $\hat{\tau}(t^{*})$ is the estimated intervention effect at exposure time $t^{*}$ using the proposed Bayesian time-varying model \eqref{eq:4_Bayes_interaction}. $b_m(.)$, $m=1,\ldots,p$, are B-spline basis functions applied to the variable exposure time $t^{*}$ and $\hat{\beta}^{*}$'s are the estimations of coefficients for the spline basis functions.

The secondary summary estimand we adopt is the intervention effect at the maximum exposure time \(t^{*}_{max} = T-2 \), representing the effect at the maximum duration of intervention. If we posit that the intervention effect trajectory plateaus after the end of the study, i.e., \( \tau(t^{*}) = \tau(T-2) \) for all \( t^{*} \geq T-2 \), then \( \Psi_{T-2} \) can be interpreted as the long-term treatment effect (LTE):
\[ \hat{\Psi}_{t^{*}_{max}}
= \hat{\tau}(T-2) = \sum_{m=1}^{p}\hat{\beta}^{*}_mb_m(T-2).
\]

In Figure \ref{fig:vary_generate}, TATE and LTE are illustrated. TATE is calculated by dividing the area under the time-varying intervention effect curve by the total length of the exposure time. This measure represents the average effect of the intervention per unit of time throughout the study's duration. LTE captures the effect of the intervention at the study's longest exposure time point, focusing on long-term outcomes. Researchers aiming to assess the intervention's average impact throughout the entire exposure period should consider TATE. On the other hand, those more interested in the sustained, long-term effects of the intervention, rather than its effect over time, would find LTE to be a more relevant estimand.

For our comparative analysis using the proposed Bayesian spline model, as detailed in Equation \eqref{eq:4_Bayes_interaction}, we evaluated the monotone effect curve model \cite{kenny2022analysis}. To the best of our knowledge, this monotone effect curve model is the only known Bayesian approach for modeling time-varying intervention effects. This model employs a monotone effect curve, assigning distinct coefficients for each study time point, while modeling the intervention effect curve over exposure time as a monotonic step function. To specifically showcase the benefits of our Bayesian penalized spline method for handling time-varying intervention effects (Model \eqref{eq:4_Bayes_interaction}), we incorporated our approach for modeling study time, as detailed in the proposed Bayesian immediate effect model \eqref{eq:Bayes}, with an adaptation of the Monotone effect curve model's monotonic step function for modeling intervention effects that change across exposure time. This focused analysis allows us to distinctly assess the impact and improvements resulting from the spline modeling of time-varying intervention effects. The Bayesian monotone effect curve model of Kenny et al.\cite{kenny2022analysis}
is defined as:
\begin{align} 
  \begin{split}
  Y_{ijt} &= \alpha  + s_j(t) + \tau(t^{*}_j) \cdot A_{jt} + \epsilon_{ijt}\\ 
  &=\alpha  + \boldsymbol{\beta}_{bj} \cdot \mathbf{B} + \delta\cdot(\sum_{\gamma=1}^{t^{*}_j}\alpha_\gamma)\cdot A_{jt} + \epsilon_{ijt}\\ 
    \epsilon_{ijt} & \sim t_\text{student}(df=3, \mu=0, \sigma=2.5)\\
\alpha & \sim \text{Normal}(0, 1) \\
\boldsymbol{\beta}_{bj} & \sim \text{Normal}(\boldsymbol{\beta}, \sigma_b^2I_{p \times p})\\
\beta_{1}& \sim \text{Normal}(0, 1)\\
\beta_{m}& \sim \text{Normal}(\beta_{m-1}, \sigma_{\beta}^2),\quad m=2,\ldots,p\\
\sigma_{b} &\sim t_\text{student}(df=3, \mu=0, \sigma=2.5)\\
\sigma_{\beta} &\sim \text{Normal}(0, 1) \\
\delta & \sim \text{Normal}(0, 5^2) \\
(\alpha_1,\ldots,\alpha_{T-2}) & \sim \text{Dirichlet}(\omega, \ldots, \omega),\\
\omega &\sim \text{Uniform}(0.01,100)
\label{eq:Bayes_monotone}
 \end{split}
\end{align} 
where $(\alpha_1,\ldots,\alpha_{T-2})$ represents a simplex where each $\alpha_{\gamma}$ is non-negative for $\gamma \in \{1,\ldots, T-2\}$. This set adheres to a symmetric Dirichlet prior, regarded as minimally informative because it does not inherently favor a significant shift or `jump' in the effect curve at any specific time point. The effect curve in this model is articulated as a monotonic step function.

In this model, the LTE estimator is simply the posterior median of $\hat{\delta}$. For the TATE, the model utilizes an estimator based on
a right-hand Riemann sum.  The process involves initially computing the posterior medians of
$(\hat{\delta}, \alpha_1, \ldots, \alpha_{T-2})$. Subsequently, we calculate the estimated intervention effect for an exposure time $t^{*}$ 
(where $t^{*}$ ranges from 1 to $T-2$). This is given by $\hat{\delta}_{t^{*}} = \hat{\delta} \cdot \sum_{\gamma=1}^{t^{*}} \alpha_{\gamma}$. 
The TATE parameter is then computed as 
$\hat{\Psi}_{(0,T-2]} = \frac{1}{T - 2} \sum_{t^{*}=1}^{T-2} \hat{\delta}_{t^{*}}$.

Our comparative analysis focuses on the bias (absolute value), RMSE, and Leave-One-Out Cross-Validation Information Criterion (LOOIC)\cite{rstanarm}.  
To compute the bias and RMSE, we used the posterior median of the parameter (TATE/LTE) as the point estimator for the parameter. LOOIC is used primarily in Bayesian model selection and evaluation, which estimates pointwise out-of-sample prediction accuracy from a fitted Bayesian model using the log-likelihood evaluated at the posterior simulations of the parameter values\cite{vehtari2017practical}. It represents expected out-of-sample deviance.
It quantifies the goodness of fit of a model while penalizing for model complexity. Models with lower LOOIC are generally preferred as they indicate a better balance between fitting the data well and not overfitting. We report the mean of LOOIC based on 150 simulations.


\begin{figure}[htbp]
  \centering
\includegraphics[width=\textwidth]{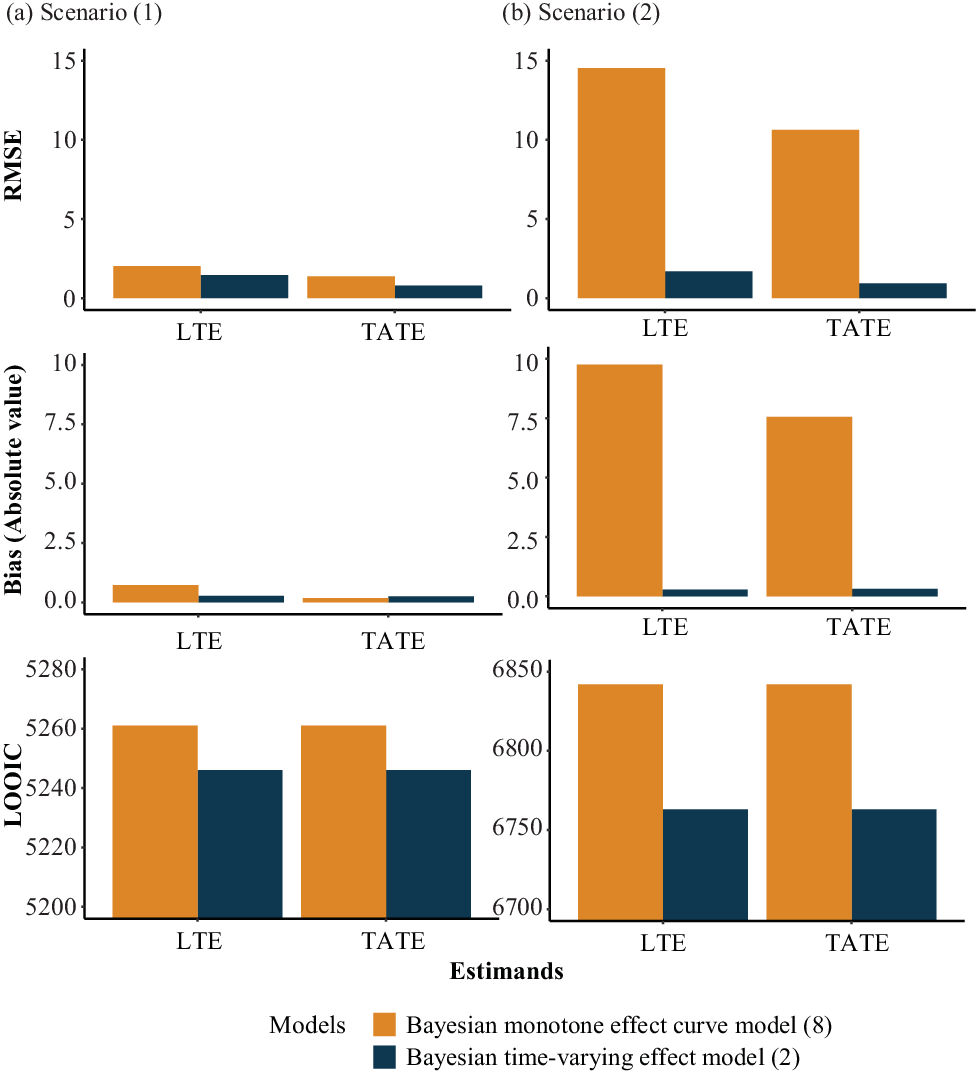}
   \caption{Comparison of the proposed Bayesian model (2) for time-varying intervention effect and the existing Bayesian monotone effect curve model with respect to bias (absolute value), RMSE, and Leave-One-Out Cross-Validation Information Criterion (LOOIC) .}
  \label{fig: Bayes_interaction_vs_monontone}
\end{figure}

Figure \ref{fig: Bayes_interaction_vs_monontone} shows the results of comparative analysis using 150 simulations. For TATE in Scenario (1), the proposed Bayesian hierarchical spline model (Model \eqref{eq:4_Bayes_interaction}) exhibits a similar bias as the monotone effect curve model (Model \eqref{eq:Bayes_monotone}). However, the proposed Model \eqref{eq:4_Bayes_interaction} has a lower RMSE of 0.79, compared to 1.38 for Model \eqref{eq:Bayes_monotone}, indicating better model accuracy. For LTE, the proposed Bayesian hierarchical spline model \eqref{eq:4_Bayes_interaction} has lower bias and RMSE, reinforcing its better estimation accuracy. Additionally, the proposed Model \eqref{eq:4_Bayes_interaction} has a lower LOOIC, indicating that it has a better fit to the data.

The results of the second scenario indicate that the proposed Bayesian hierarchical spline model (Model \eqref{eq:4_Bayes_interaction}) clearly outperforms the monotone effect curve model (Model \eqref{eq:Bayes_monotone}) in terms of bias (absolute value), RMSE and LOOIC for both TATE and LTE. For TATE, the proposed Bayesian model \eqref{eq:4_Bayes_interaction} exhibits a  minimal bias at 0.31 compared to 7.55 seen in the existing monotone effect curve model \eqref{eq:Bayes_monotone}. Additionally, the RMSE is improved, with the proposed Model \eqref{eq:4_Bayes_interaction} showing an RMSE of 0.93, which is significantly lower than the RMSE of Model \eqref{eq:Bayes_monotone} at 10.63. Similarly, for LTE, the proposed Model \eqref{eq:4_Bayes_interaction} demonstrates its superiority with a lower bias of 0.28 against the existing Model \eqref{eq:Bayes_monotone}'s 9.75. The RMSE for the proposed model \eqref{eq:4_Bayes_interaction} stands at 1.70, which is significantly lower than the existing Model \eqref{eq:Bayes_monotone}'s RMSE of 14.53. We also compared Model \eqref{eq:4_Bayes_interaction} with Model \eqref{eq:Bayes_monotone} in terms of the R-hat diagnostic, which is crucial for assessing the convergence of chains in Hamiltonian Monte Carlo (HMC) simulations \cite{StanDevelopmentTeam2020}. The proposed Model \eqref{eq:4_Bayes_interaction} can achieve model convergence with a shorter chain, thus saving computational resources.

These results underscore the benefits of using the proposed Bayesian spline Model \eqref{eq:4_Bayes_interaction}, highlighting its increased accuracy, reliability, and flexibility in estimating time-varying intervention effects in SWCRTs, making it a valuable tool for researchers and practitioners seeking to understand and evaluate the impacts of interventions accurately.

\subsubsection{Performance evaluation for Bayesian cluster-specific time varying intervention effect model}

In this section, we present the simulations conducted to assess the performance of the proposed Bayesian spline Model \eqref{eq:4_Bayes_interaction_randomeff}, which incorporates cluster-specific random effect for time-varying intervention effects.  This model can accommodate variations in intervention effects across different clusters. The simulation setup is based on Scenario (3) where the intervention is applied across $J = 10$ clusters, each cluster contains 10 participants ($I=10$), and the number of time periods $T_{scenario 3} = 22$. To generate the intervention effect curve for each cluster, we used the following setting. 

\begin{itemize}
    \item Scenario (3):
    \begin{itemize}
    \item Overall time-varying intervention effect curve
        \[ \tau(t^{*}_j) = 
\begin{cases} 
0 & t^{*}_j=0\\
\frac{5}{1 + e^{-5 \times (t - 1)}} & \text{if } 5 > t^{*}_j > 0 \\
2.5 + \frac{5}{1 + e^{2 \times (t - 5)}} & \text{if } t^{*}_j \geq 5 
\end{cases}
\]
\item Cluster-specific intervention effect curves
\[
 \tau_j(t^{*}_j) = \tau(t^{*}_j) \cdot exp(u_j),\quad
u_j \sim \text{Normal}(0,0.2^2),\quad j\in\{1,\dots,J\}
\]
\end{itemize}
\end{itemize}

\begin{figure}[h]
  \centering
\includegraphics{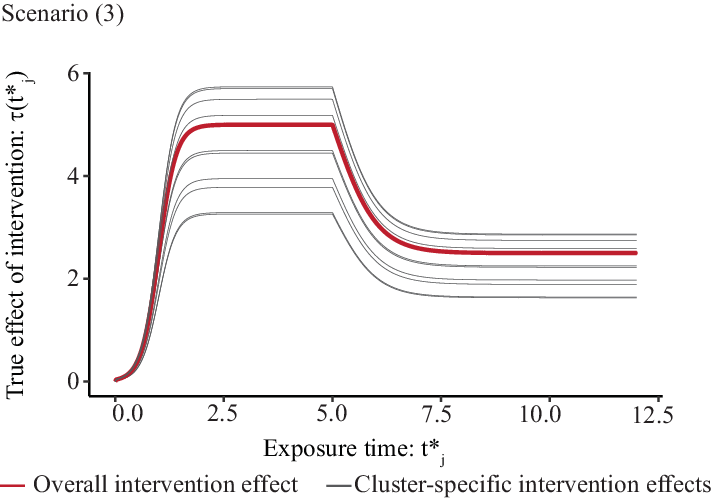}
   \caption{Scenario of generating cluster-specific time-varying intervention effects. }
  \label{fig:cluster_specific_t_vary_generation}
\end{figure}
\clearpage

\begin{figure}[p]
  \centering
\includegraphics{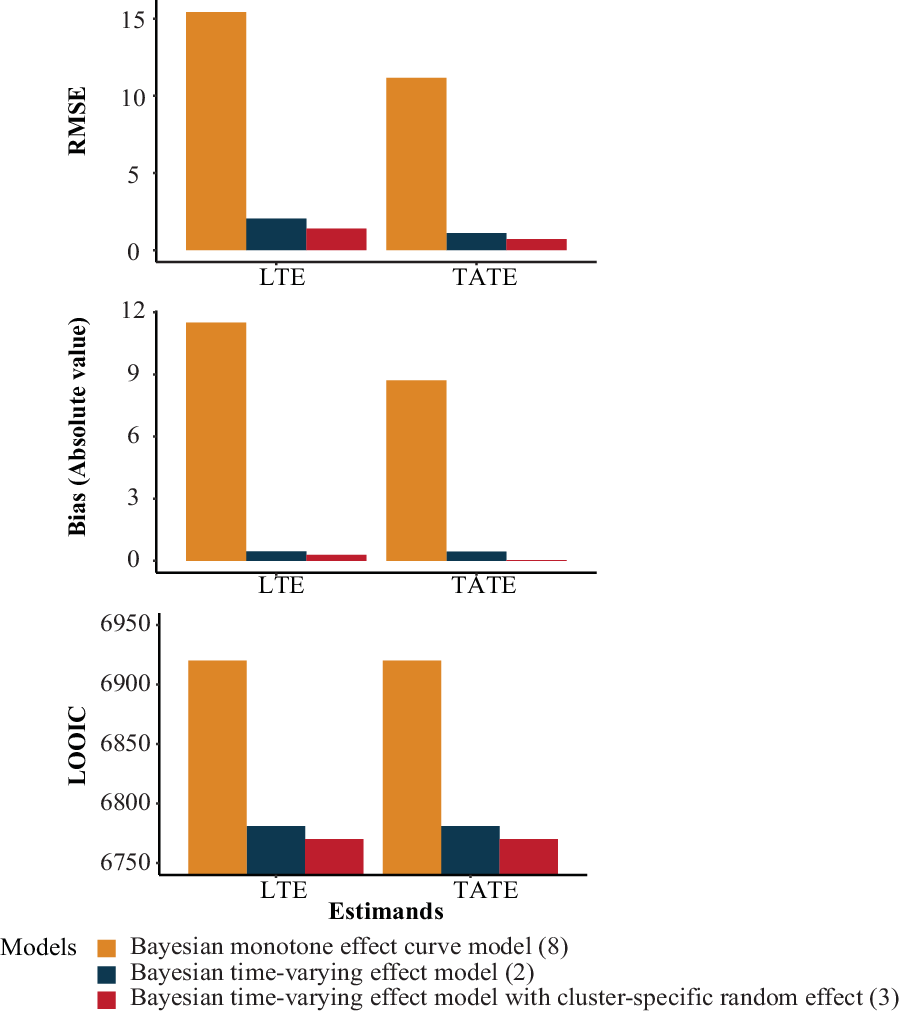}
   \caption{Comparative performance of the Bayesian cluster-specific time-varying effect model (Model \eqref{eq:4_Bayes_interaction_randomeff}), the Bayesian time-varying effect model (Model \eqref{eq:4_Bayes_interaction}), and the existing Bayesian monotone effect curve model (Model \eqref{eq:Bayes_monotone}) in terms of RMSE, bias (absolute value), and Leave-One-Out Cross-Validation Information Criterion (LOOIC; Expected Out-of-Sample Deviance)}
  \label{fig:cluster_specific_t_vary_performance}
\end{figure}

Figure \ref{fig:cluster_specific_t_vary_generation} shows the true value of the intervention effects from one simulation. We conducted 150 simulations, each randomly generating $u_j$ from Normal($0,0.2^2$) while maintaining the same overall intervention effect curve across all simulations. We conducted a comparative analysis between the proposed Bayesian cluster-specific time-varying effect model \eqref{eq:4_Bayes_interaction_randomeff} and the proposed Bayesian time-varying effect model \eqref{eq:4_Bayes_interaction} (no cluster-specific random effect for the intervention effects) and the existing Bayesian monotone effect curve model \eqref{eq:Bayes_monotone}. This comparison was focused on evaluating their performances based on RMSE, bias (absolute value), and LOOIC using 150 simulations. The results are illustrated in Figure \ref{fig:cluster_specific_t_vary_performance}.

The results demonstrate the superior performance of the Bayesian model \eqref{eq:4_Bayes_interaction_randomeff} across both TATE and LTE estimands. For the LTE estimand, the proposed Bayesian cluster-specific time-varying effect model (Model \eqref{eq:4_Bayes_interaction_randomeff})  achieved the lowest RMSE at 1.39, and a significantly lower bias of 0.29, indicating its precision in estimation. Similarly, for the TATE estimand, Model \eqref{eq:4_Bayes_interaction_randomeff}  outperformed others with the lowest RMSE of 0.71 and a minimal bias of 0.03. Additionally, the mean of LOOIC values, computed from 150 simulations, for the proposed Bayesian cluster-specific time-varying effect model (Model \eqref{eq:4_Bayes_interaction_randomeff})  was the lowest.

The performance of the Bayesian cluster-specific time-varying model \eqref{eq:4_Bayes_interaction_randomeff} is evaluated by examining the entire curves of intervention effect and temporal trends. Appendix A.4.1 presents the posterior distributions for the intervention effects at each exposure time point alongside the actual true values. Appendix A.4.2, on the other hand, displays the posterior distributions for the temporal trend curve at each point in the study period, in comparison to the true values. These results visually demonstrate the model's estimation accuracy in estimating the time-varying intervention effect and temporal trends.

\subsection{Application to date from a stepped wedge cluster-randomized trial}
\label{sec:4_application}

We apply the proposed Bayesian hierarchical spline model to analyze the ``Primary Palliative Care for Emergency Medicine (PRIM-ER)" study \cite{grudzen2019primary}, a stepped wedge cluster-randomized trial aimed at evaluating the effectiveness of primary palliative care education, training, and technical support in emergency medicine. The study, analyzed across 29 emergency departments (EDs) in the United States, includes about 100,000 older patients with serious, life-limiting illnesses. The intervention is implemented in a stepped manner, with new ED sites adopting the PRIM-ER intervention every three weeks over a 2-year period. This intervention comprises four components, which were designed to increase primary palliative care skills and knowledge among emergency providers, thereby aligning care plans more closely with patients' health goals and improving patient-centred outcomes:

\begin{itemize}
    \item Evidence-Based Multidisciplinary Primary Palliative Care Education: full-time emergency physicians, nurse practitioners, and physician assistants engage in online didactic courses and simulation-based workshops, focusing on communication in serious illness.
    \item Simulation-Based Workshops: practical, hands-on training in end-of-life communication within a controlled environment.
    \item Clinical Decision Support: system that integrates alerts and new workflows into electronic health records to facilitate referrals to palliative care, home care, and hospice services.
    \item Audit and Feedback: emergency providers receive individualized reports to track and improve their number of referrals to palliative care, home care, and hospice services.
\end{itemize}

The primary hypothesis is that the intervention will lead to a reduction in ED disposition
to an acute care setting for older patients with serious, life-limiting illnesses who visit the ED for any reason. The primary outcome assessed is the proportion of eligible
patients whose disposition is to an acute care setting (inpatient,
non-palliative service) treated as a dichotomous variable (yes/no for acute care admission).

\begin{figure}[p]
  \centering
 \includegraphics[width=\textwidth]{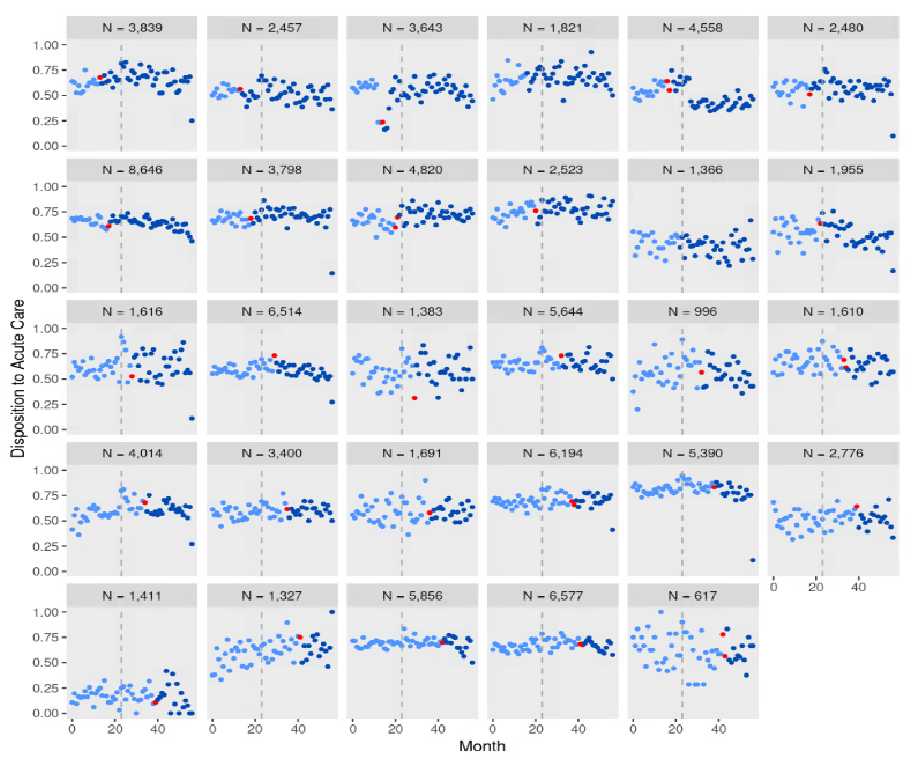}
   \caption{The observed proportions of eligible patients who were discharged to an acute care setting. The vertical lines indicate the start of the pandemic. The light blue points indicate the proportion of acute care admissions in the control state, the red points indicate the proportion during the 3-week transition period (when training was happening), and the dark blue points indicate the proportion once the cluster was using the intervention.}
  \label{fig:primer}
\end{figure}

Figure \ref{fig:primer} shows the observed proportions of eligible patients who were discharged to an acute care setting in the PRIM-ER study. The PRIME-ER intervention was implemented in the first ED in May 2019 and the final ED transitioned to the intervention in November 2021. The vertical lines indicate the start of the pandemic. The light blue points indicate the proportion of acute care admissions in the control state, the red points indicate the proportion during the 3-week transition period (when training was happening), and the dark blue points indicate the proportion once the cluster was using the intervention. Because the COVID-19 pandemic largely played out in EDs across the country during the study time, the effect of the PRIM-ER ED intervention may have been different before and during the pandemic. To account for
this, the analyses will include pandemic-period effects for both the control and intervention states to account
for differences in outcome and possibly treatment effects as a result of the pandemic. We fit the proposed Bayesian spline models (i.e., the Bayesian immediate effect model \eqref{eq:Bayes} and the Bayesian cluster-specific time-varying effect model \eqref{eq:4_Bayes_interaction_randomeff} ) with adjustment of the COVID-19 pandemic period (Yes/No) as well as an interaction term between the COVID-19 pandemic period (Yes/No) and Intervention states (Intervention/Control condition).

\clearpage


\begin{figure}[p]
  \centering
 \includegraphics[width=\textwidth]{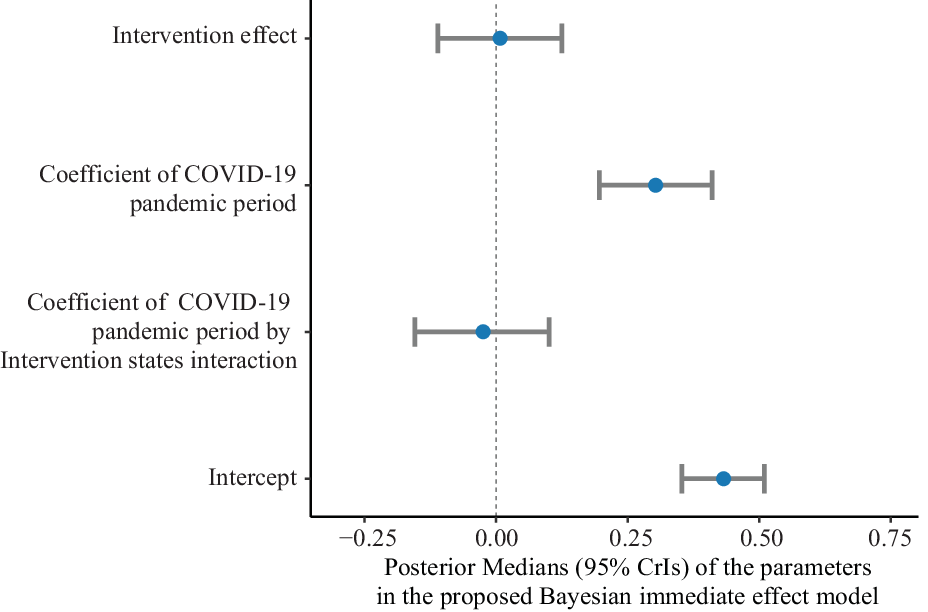}
   \caption{The proposed Bayesian immediate effect model (1) applied to the PRIM-ER Study: a summary of model parameters' posterior distributions including posterior median and 95\% credible interval.}
  \label{fig:primer_immed_model}
\end{figure}
\clearpage

\begin{figure}[p]
  \centering
 \includegraphics[width=\textwidth]{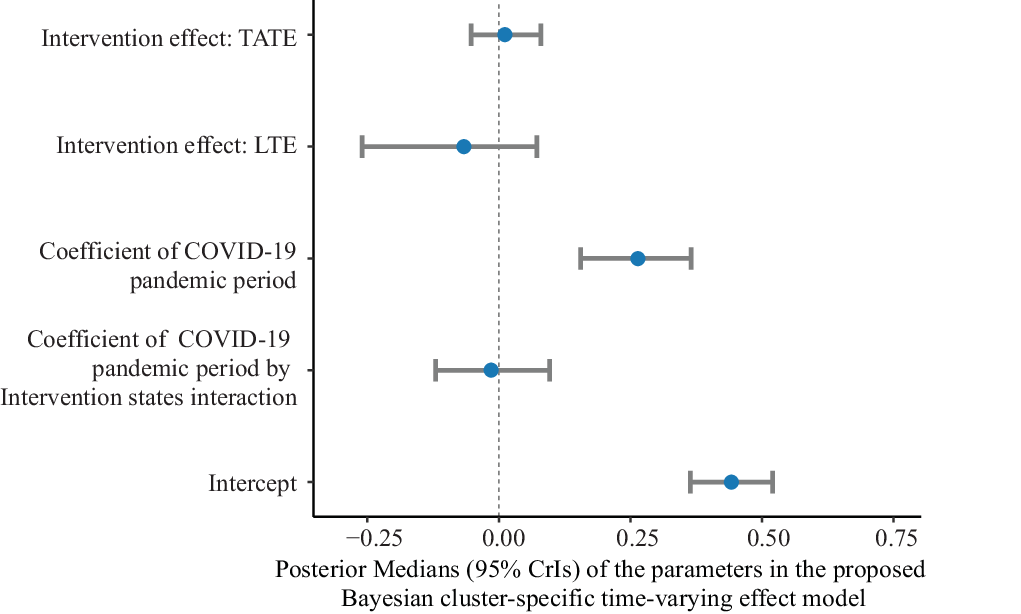}
   \caption{The proposed Bayesian cluster-specific time-varying effect model (3) applied to the PRIM-ER Study: a summary of posterior distributions of TATE and LTE (two estimands for summarizing the magnitude of the overall intervention effect curve), along with other regression coefficients}
  \label{fig:primer_timevary_model}
\end{figure}
\clearpage

Forest plot \ref{fig:primer_immed_model} shows the results of fitting the proposed Bayesian immediate effect model \eqref{eq:Bayes} to the data from the PRIM-ER study, the posterior distribution of the intervention effect ($\tau$) indicates that the intervention does not significantly lead to a reduction in the number of patients being admitted to acute care settings from the ED. Forest plot \ref{fig:primer_timevary_model} shows the results of fitting the proposed Bayesian cluster-specific time-varying effect model \eqref{eq:4_Bayes_interaction_randomeff} to the data from the PRIM-ER study. The two estimands for summarizing the intervention effect curve, TATE and LTE, both of which include 0 within their credible intervals, suggest the same conclusion from the Bayesian immediate effect model \eqref{eq:Bayes}: a null effect of the intervention, contrary to the primary hypothesis. This suggests that the implementation of the PRIM-ER components, designed to enhance primary palliative care skills among emergency providers and align care plans with patients' health goals, did not significantly change the rate of acute care admissions. These findings prompt a reevaluation of the intervention's strategies and raise questions about the factors that might have influenced this outcome. It underscores the complexity of healthcare interventions in emergency settings and the need for a deeper understanding of how such interventions interact with varied patient populations and healthcare systems.

\section{Discussion and conclusions}
\label{sec:discuss}
The presented work advances the Bayesian methodologies in the field of SWCRTs by introducing Bayesian hierarchical penalized spline models. These models adeptly address the pivotal challenges of managing time-related confounding and accurately estimating both immediate and time-varying intervention effects. The proposed Bayesian immediate effect model \eqref{eq:Bayes} is designed for SWCRTs with a relatively large number of clusters and time periods, focusing on immediate intervention effects. Demonstrated through extensive simulations, this model consistently attains nearly frequentist nominal coverage probability for the intervention's true effect, solving the low coverage problem using conventional frequentist methods and offering a robust statistical alternative.
 
 The proposed Bayesian time-varying effect model \eqref{eq:4_Bayes_interaction}, which accommodates time-varying intervention effects, utilizes the versatility of Bayesian penalized splines to model how treatment effects vary with intricate temporal trends. The precision of this model is highlighted by its superior performance in bias and RMSE measures compared to the existing Bayesian monotone effect curve model (model \eqref{eq:Bayes_monotone}). Meanwhile, the proposed Bayesian spline model \eqref{eq:4_Bayes_interaction} provides better reliability and stability of the results obtained from HMC simulations in Bayesian modeling, as indicated by LOOIC and R-hat, thus advancing Bayesian methodology in SWCRTs. 

Our work should be interpreted in the context of a potential limitation. The adoption of Bayesian approaches has increased significantly due to the advancements in modern computing power and the efficiency of MCMC algorithms \cite{Berry2010BayesianTrials}. However, our proposed Bayesian methods may still require a higher computational overhead compared to the frequentist GAM approach. In SWCRTs with sample sizes ranging from several hundreds to thousands, the computational time of the proposed Bayesian methods remains comparable to that of the frequentist GAM method. However, in scenarios involving a large number of subjects, for example, 900,000, the difference becomes stark: the GAM methods can complete in one minute, whereas the proposed Bayesian immediate effect model \eqref{eq:Bayes} requires approximately 15 hours. Despite the longer computation time, the proposed Bayesian spline models offer higher accuracy and better coverage, as illustrated by our simulations.  Given these benefits, future research can focus on developing ways to expedite the Bayesian estimation process. Such advancements would not only improve the practicality of Bayesian modeling in large-scale studies but also ensure a balanced approach between computational expediency and analytical accuracy. 

To the best of our knowledge, the development of the Bayesian penalized spline model has not been undertaken previously in SWCRTs. In conducting this study, we believe we have developed a versatile tool that can be used in future research to model the time trend and immediate and time-varying intervention effects for SWCRTs. This adaptability emphasizes the model's significance as a contribution to Bayesian modeling, promising to enhance future research and analysis within stepped wedge designs.

\section{Acknowledgement}
We thank Nina Siman, MA, MSEd (NYU Grossman School of Medicine) for her critical role in developing the PRIM-ER analysis data set and files. We thank Dr. Eva Petkova for her support. Research reported in this publication was supported within the National Institutes of Health (NIH) Health Care Systems Research Collaboratory by cooperative agreement UG3AT009844 from the National Center for Complementary and Integrative Health, and the National Institute on Aging. This work also received logistical and technical support from the NIH Collaboratory Coordinating Center through cooperative agreement U24AT009676. The content is solely the responsibility of the authors and does not necessarily represent the official views of the National Institutes of Health.
\section{Code}
Code is available: https://github.com/dannidanniwu/BSW/Final simulation code


\clearpage

\section{Appendices}
A.1 \:  Assessing the estimation accuracy of time-varying intervention effects using the proposed Bayesian cluster-specific time-varying model \eqref{eq:4_Bayes_interaction_randomeff}

\begin{figure}[h]
\renewcommand\thefigure{A1}
\centering
\includegraphics[]{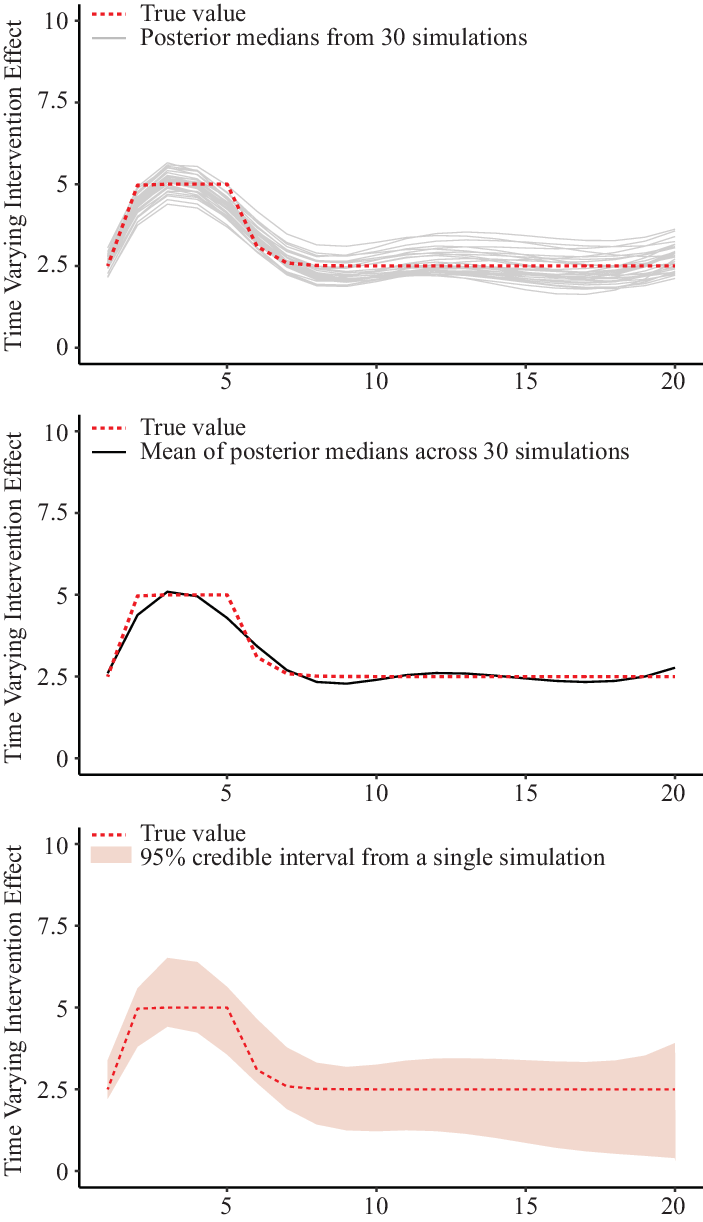}
\caption{Assessing the estimation accuracy of time-varying intervention effects using the proposed Bayesian cluster-specific time-varying model \eqref{eq:4_Bayes_interaction_randomeff}}
\label{fig:time_vary_trt_vs_true}
\end{figure}

Figure \ref{fig:time_vary_trt_vs_true} presents three panels illustrating the performance of the proposed Bayesian cluster-specific time-varying model \eqref{eq:4_Bayes_interaction_randomeff} in estimating time-varying intervention effects through simulations. The top panel displays posterior medians for the intervention effect at each exposure time point from 30 simulations, with the true value overlaid in red. The middle panel shows the mean of these posterior medians across the 30 simulations compared to the true value. The bottom panel presents the 95\% CrI for the intervention effect from a randomly selected single simulation, with the true intervention effect curve plotted for reference. The width of the credible interval suggests the degree of uncertainty in the model's estimates at different time points. Overall, these panels collectively demonstrate the model's ability to capture the true intervention effect with a reasonable degree of accuracy and precision.

\clearpage

A.2 \:  Assessing the estimation accuracy of temporal trends using the proposed Bayesian cluster-specific time-varying model \eqref{eq:4_Bayes_interaction_randomeff}

\begin{figure}[h]
\renewcommand\thefigure{A2}
\centering
\includegraphics[]{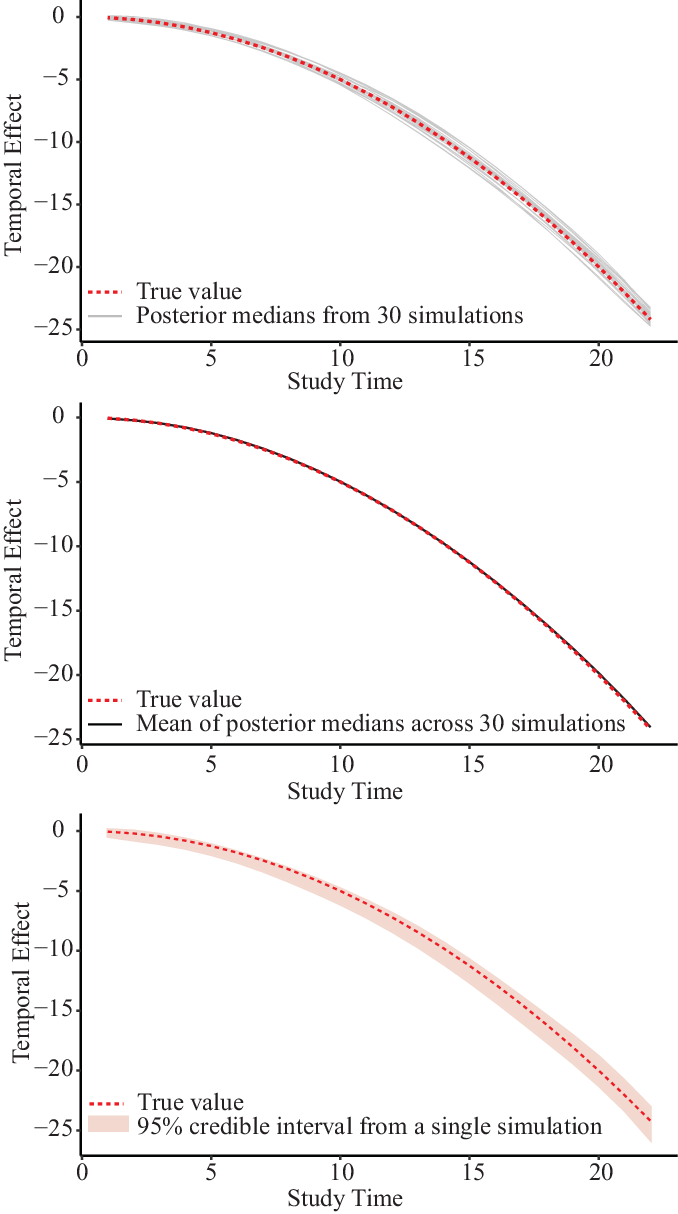}
\caption{Assessing the estimation accuracy of temporal trends using the proposed Bayesian cluster-specific time-varying model \eqref{eq:4_Bayes_interaction_randomeff}.}
\label{fig:time_effect_vs_true}
\end{figure}

Figure \ref{fig:time_effect_vs_true} presents the estimated temporal effect over a study period by the proposed Bayesian cluster-specific time-varying model \eqref{eq:4_Bayes_interaction_randomeff} using simulated data. The top panel illustrates the posterior medians of the temporal effect at each study time point estimated from 30 simulations, plotted against the true temporal effect. These posterior medians closely align with the red dotted line representing the true value, indicating a consistent model performance across simulations. The middle panel displays the mean of the posterior medians from the same 30 simulations juxtaposed with the true value. Finally, the bottom panel focuses on uncertainty quantification, showing the 95\% CrI of the temporal effect from a randomly selected single simulation, with the true temporal effect overlaid for reference. These plots underscore the model's capability to not only accurately estimate the true temporal effect but also quantify the inherent uncertainty of such estimates over time.

\end{document}